\documentstyle[12pt,lscape]{article}
\begin{document}
\title{\vspace*{-1.8cm}
Charming penguins in $B \to PP $ decays
and the extraction of $\gamma $}
\author{
{P. \.Zenczykowski$^*$ }\\
\\
{\em Dept. of Theoretical Physics},
{\em Institute of Nuclear Physics}\\
{\em Polish Academy of Sciences}\\
{\em Radzikowskiego 152,
31-342 Krak\'ow, Poland}\\
}
\maketitle
\begin{abstract}
It is shown that inclusion of charming penguins of the size
suggested by short-distance dynamics may shift down by
$10^o-15^o$ the value 
of $\gamma $ extracted via the overall fit to the
$B \to PP $ branching ratios. A substantial dependence of the fit 
on their precise values is found,  
underscoring the need to improve the reliability of
data.

\end{abstract}
$^*$ E-mail:
zenczyko@iblis.ifj.edu.pl
\newpage

\section{Introduction}
Various methods of extracting the
value of the unitarity-triangle angle $\gamma $ 
from data have been proposed in the literature. 
 Some of them are based on the analysis
of the decays of $B$ mesons into a pair of light 
pseudoscalar mesons $PP$, and in particular into the $K\pi$ states. 
With most present data on asymmetries in $B \to PP$ decays still carrying large
errors, fits to the branching ratios and asymmetries of $B \to PP$ decays
depend mainly on the former.

In the simplest approach \cite{Rosner1994} to these decays
the full $B \to PP$ amplitudes are given
in terms of only a few short-distance (SD)
amplitudes corresponding to specific quark-line diagrams 
(tree $T$, colour-suppressed $C$,
penguin $P$, singlet penguin $S$) expected to provide the dominant
contributions. The penguin amplitude is furthermore assumed to be
dominated by the contribution from the internal top quark propagation
\cite{Rosnerconventions}.
The only electroweak penguin that has to be kept is included
through an appropriate replacement in colour-suppressed strangeness-changing
amplitude. The value of angle $\gamma $ 
extracted from such analyses depends of course
on strong SD phases and on possible modifications
of the SD formulae by aditional effects.
Among the latter effects the issue of the size of rescattering contribution
has been addressed by several investigators. 

The rescattering (or final state interaction - FSI) contribution
is composed of two main parts: the contribution in which the intermediate
state contains charmed quarks (so-called charming penguins) 
\cite{BurasFleischer1995,Ciuchini,BurasSilvestrini,Pham,Rosner2001,GerardSmith2003}, and the
contribution from elastic and inelastic rescattering through intermediate
states involving only light (ie. $u$, $d$, $s$) quarks
\cite{Wolf,FSI,Hou,Zen2001,otherFSI,LZ2002}. 
In a recent paper \cite{ZL2003} the latter contribution was analysed
in detail for the SU(3)-symmetry breaking case.
The main conclusion of ref \cite{ZL2003} was that inclusion of such
effects may significantly affect the extracted value of angle $\gamma $.
Namely, while for negligible strong SD phases the global fits to 
the branching ratios of all $B \to PP $ decays yielded the value of 
$\gamma \approx 100^o$, similar fits with rescattering effects 
included permitted values of $\gamma $ in a broad range of $(50^o,110^o)$,
and actually even preferred a value of $\gamma $ in 
qualitative agreement
with SM expectations of $\gamma _{SM} \approx 65^o $. 

Paper \cite{ZL2003} left open the issue of the effect induced by
charming penguins.
Furthermore, the size of the contribution from inelastic rescattering,
required for the 
shift of the extracted value of $\gamma$ down 
by some $30^o$,   was a factor of five larger
than the estimates of the size of quasi-elastic rescattering in a Regge model
\cite{Zen2001}.
Given low experimental bounds on the size of the observed branching ratios of 
the $B \to K\bar{K}$ decays, which are thought to provide a bound on the
size of rescattering effects,
one might therefore argue that these effects should be much smaller than those
resulting from the fits of ref.\cite{ZL2003}.

In the present paper we address again the question of the size of 
corrections to the dominant $t$-quark contributions to penguin amplitudes,
and show that shifts in the extracted value of $\gamma $
of the order of $10^o-15^o$ may result from the inclusion of SD
charming penguins. 
Furthermore, we observe that the use of the updated values of the $B \to PP$  
branching ratios shifts the value of $\gamma $ extracted 
when no rescattering is considered
down by $20^o$ when compared to the fit of 
\cite{ZL2003} . Although for the recent values
of branching ratios the agreement with the data is now worse than in
ref.\cite{ZL2003}, the data do point out to a lower
value of $\gamma $.

\section{Dominant short-distance amplitudes}

In this paper  the dominant short-distance amplitudes
are parametrized exactly as in \cite{ZL2003}. Thus, we assume that all their
strong phases are negligible. 
Although these phases may be non-zero \cite{perturb,PQCD},
their precise values are not relevant for what we want to discuss here:
the aim of this paper is to look at uncertainties not related to
these phases (as long as the latter remain small).

Thus, for the tree amplitudes we use
\begin{equation}
T'=\frac{V_{us}}{V_{ud}} \frac{f_K}{f_{\pi}} T \approx 0.276 ~T 
\end{equation}
with (un)/primed amplitudes denoting strangeness (preserving)/changing
processes.
Both tree amplitudes have the same weak phase: 
$T/|T|=T'/|T'|=e^{i \gamma }$.

Assuming that the penguin SD amplitudes are dominated by the $t$ quark, 
the weak phase factor is $e^{-i \beta}$ for $P$ and $- 1$ for $P'$
(ie. $P'= - |P'|$). 
We use the estimate \cite{CR2001}
\begin{equation}
\label{penguins}
P=-e^{-i \beta}\left | \frac{V_{td}}{V_{ts}}\right | P'
\approx -0.176~ e^{-i \beta }P'.
\end{equation} 
In the following we use $\beta \approx 24^o$,
which is in agreement with the world average \cite{beta}
$\sin 2\beta = 0.734 \pm 0.054$. 

We accept the relations between
the tree and the colour-suppressed amplitudes given by the SD estimates:
\begin{equation}
C = \xi T
\end{equation}
and
\begin{equation}
\label{cprime}
C'=T'(\xi -(1+\xi) \delta _{EW} e^{-i\gamma})
\end{equation}
where we take $\xi =\frac{C_1+\zeta C_2}{C_2+\zeta C_1}\approx 0.17 $,
assuming $\zeta \approx 0.42$, ie. midway between $1/N_c$ and the value of
$0.5$ suggested by experiment, and using $C_1\approx -0.31$ and $C_2 \approx 
1.14$ \cite{BBurasL}.
The contribution from the electroweak penguin $P'_{EW}$ 
has been included in Eq.(\ref{cprime}),
with $\delta _{EW}\approx +0.65$ \cite{EWP} (other
electroweak penguins are neglected).

Finally, since data suggests that the singlet penguin
amplitude $S'$ is sizable 
(cf. \cite{DGR97PRL79,CR2001}) we include it in our calculations
as well, with weak and strong phases as for $P'$.
The remaining SD amplitudes (exchange $E$ and $E'$, 
singlet penguin $S$, penguin
annihilation $PA$ etc.) are neglected.
Thus, the dominant SD amplitudes depend on four SD parameters: 
$|T|$, $P'$, $S'$, and the weak phase $\gamma $.

Because rescattering effects induced by Pomeron exchange
are fully calculable, we correct for them
following ref.\cite{ZL2003} (the relevant theoretical formulae for
all $B \to PP$ amplitudes in question are given
in Table 1 therein). Actually, it is
only when SU(3) is broken that these corrections are different for
different decay channels, and the resulting deviations from 
the standard SD form could be observed.

As in ref.\cite{ZL2003} we minimize the $\chi ^2$ function
defined as:
\begin{equation}
\label{chi2}
\chi ^2 = \sum _i \frac{({\cal B}^{the}_i-{\cal B}^{exp}_i)^2}
{(\Delta {\cal B}_i)^2}
\end{equation}
where ${\cal B}^{the(exp)}_i$ denote theoretical (experimental) 
CP-averaged
branching ratio for the $i$-th decay channel. We consider the same 
16 decay channels as in ref.\cite{ZL2003} (see Table 1). 
Their experimental
branching ratios and errors
taken from \cite{HFAG2003}
are given in the second column of Table 1. 
These numbers differ from the ones used in \cite{ZL2003} in a couple of
entries, the most significant ones (ie. where the new average is more than one
old standard deviation away from the old average) being for $\pi ^+ \eta$, 
$\pi ^+ K^0$, and $\pi ^0 K^0$.
In the calculations
themselves, the branching ratios were corrected for the deviation
of the ratio of the $\tau _{B^+}$ and $\tau _{B_0}$ lifetimes from unity 
(using $\tau _{B^+}/\tau _{B_0}=1.086 $).
For a given value of $\gamma $ the $\chi ^2$ function 
was minimized with respect to $|T|$, $P'$, and $S'$.

The resulting dependence on $\gamma $ is shown in Fig.1 as solid line.
The fitted values of the branching ratios  together
with their deviations from the experimental numbers are given in columns 3 and 4
of Table 1.
When comparing with the fits of ref.\cite{ZL2003} one observes a strong
shift of the minimum (from just above $100^o$ in \cite{ZL2003} to $82^o$
here), and a significant increase in the size of $\chi ^2$ (from $14.3$
to $26.0$). The size of both shifts underscores the need to improve
the reliability of data. One observes that the updated fit has problems with
the description of not only $\pi ^0 \pi ^0$ and $\pi ^0 K^0$  
as in ref.\cite{ZL2003}, but also, though to a lesser extent,
 with $\pi ^+ \pi ^0$ and $\pi ^+ \eta $.

\section{Rescattering effects and short-distance charming penguins}
The fits of the preceding section assumed SD penguin amplitudes
to be totally dominated by top quark contribution $P_t$.
Various kinds of rescattering effects generate additional
contributions due to intermediate
$u$ and $c$ quarks, and may modify $P_t$ 
so that
the full penguin contributions 
(denoted by $\tilde{\phantom{xx}}$ ) may be written as:
\begin{eqnarray}
\label{Puct}
\tilde{P}&=&\lambda ^{(d)}_u\tilde{P}_u+\lambda ^{(d)}_c\tilde{P}_c+
\lambda ^{(d)}_t\tilde{P}_t \\
\label{Puctprim}
\tilde{P}'&=&\lambda ^{(s)}_u\tilde{P}_u+\lambda ^{(s)}_c\tilde{P}_c+
\lambda ^{(s)}_t\tilde{P}_t
\end{eqnarray}
where
\begin{equation}
\lambda ^{(k)}_q=V_{qk}V^*_{qb},
\end{equation}
with $V$ being the CKM matrix.

Ref.\cite{ZL2003} was concerned with contributions of $\tilde{P}_u$ type. In the
SU(3)-symmetry breaking case studied in \cite{ZL2003} this
contribution varied
from channel to channel. Its SU(3)-symmetric part 
was parametrized by a single complex 
 parameter $d$ (one of three effective FSI parameters discussed in
\cite{ZL2003}), 
so that for SU(3)-symmetric FSIs
all formulas for individual $B \to PP$ strangeness-changing
amplitudes in \cite{ZL2003} depended
on a single 
FSI-corrected penguin amplitude:
\begin{equation}
\tilde{P'}= P'_{SD}(1+i 3d)+i d T'_{SD}
\end{equation}
where $P'_{SD}=\lambda ^{(s)}_tP_t$, and $T'_{SD}=T'\propto \lambda ^{(s)}_u$.
The expression for $\tilde{P}$ was, of course, completely analogous.
It was the $i d T'$ term above which generated 
the $ \lambda ^{(s)}_u\tilde{P}_u$-type
term of Eq.(\ref{Puctprim}) in \cite{ZL2003}.
Thus, in the case of SU(3)-symmetric FSIs all rescattering effects not involving
intermediate charmed quarks can be
hidden into the $\lambda ^{(k)}_u\tilde{P}_u$ term in Eqs.(\ref{Puct},\ref{Puctprim}). 
(However,
this cannot be done in a decay-channel-independent manner if FSI break SU(3),
the case considered in \cite{ZL2003}.) 

As discussed in \cite{ZL2003}, FSI effects may depend on two further
effective parameters ($c$ and $u$). 
The first of them ($c$) takes care of "crossed"
quark-line diagrams and modifies the effective "tree" and "color-suppressed"
diagrams.
In refs.\cite{LZ2002,ZL2003} it was shown that nonzero value of $c$ leads to
effective $\tilde{T}^{^{(}{'}{^)}}$ ($\tilde{C}^{^{(}{'}{^)}}$) amplitudes being
mixtures of SD tree and colour-suppressed amplitudes with different
strong phases. The penguin and singlet penguin get similarly mixed.
Since in the fits of \cite{ZL2003} small values of $c$ were obtained,
we shall not be interested here in these corrections.
Nonzero value of the other parameter ($u$) leads to effective annihilation $A$, 
exchange $E$ and penguin annihilation $PA$ amplitudes. 
Parameters $u$ and $d$ describe the contributions from quasi-two-body
intermediate states in which the two intermediate mesons belong
to multiplets classified by the same or 
different charge conjugation parities $C$ \cite{LZ2002}.
If only states composed of two pseudoscalar mesons contributed to 
the FSI effects,
the parameters $u$ and $d$ would be proportional to each other 
($u=d/2$ in the normalization of \cite{LZ2002,ZL2003}). 
Then, from the size of $A$, $E$, $PA$ amplitudes from eg. $B^0_d \to K^+K^-$
 one could determine $u$ and evaluate the size of rescattering contribution 
 to penguin amplitudes.
 However, intermediate states of C parity opposite to that of the PP state may
 also contribute. The relation between $u$ and $d$ is then relaxed, and
 one may have small $B^o_d \to K^+K^-$ branching ratio and substantial
 FSI contribution to penguin amplitudes.
 Ref.\cite{ZL2003} was concerned with this possibility.
 The fits performed in \cite{ZL2003} suggest that the $\tilde{P}_u$ term could
 be substantial. Since a large size of this term may be questioned
 it would be worthwhile to find other arguments that could support one of the
 claims of ref.\cite{ZL2003}, namely that keeping only the $\tilde{P}_t$ 
 term may lead to a significant error in the
 extracted value of $\gamma $.
 We shall do that below on the example of the SD charming penguin.
 
 Using the unitarity property of the CKM matrix one may rewrite expressions
 (\ref{Puct},\ref{Puctprim}) as \cite{BurasFleischer1995}:
 \begin{eqnarray}
 \tilde{P}&=&\lambda ^{(d)}_c(\tilde{P}_c-\tilde{P}_u)+
 \lambda ^{(d)}_t(\tilde{P}_t-\tilde{P}_u) \\
 \tilde{P}'&=&\lambda ^{(s)}_c(\tilde{P}_c-\tilde{P}_u)+
 \lambda ^{(s)}_t(\tilde{P}_t-\tilde{P}_u) 
 \end{eqnarray}
 Since 
 \begin{eqnarray}
 \lambda ^{(d)}_t&=&
 -\lambda ^{(s)}_t \left| \frac{V_{td}}{V_{ts}} \right| e^{-{i\beta}}\\
 \lambda ^{(d)}_c&\approx & \lambda ^{(s)}_t \lambda \\
 \lambda ^{(s)}_c &\approx & -\lambda ^{(s)}_t,
 \end{eqnarray}
 where $\lambda \approx 0.22$ is the Wolfenstein parameter,
 for negligible $\tilde{P}_u$
 the above formulae may be rewritten as
 \begin{eqnarray}
 \tilde{P}&=& -\lambda ^{(s)}_t \tilde{P}_t 
 \left( \left| \frac{V_{td}}{V_{ts}} \right| e^{-{i\beta}}
 -\lambda \zeta \right)\\
 \label{PSprim}
 \tilde{P}'&=&\lambda ^{(s)}_t \tilde{P}_t(1-\zeta)
 \end{eqnarray}
 with
 \begin{equation}
 \zeta = \frac{\tilde{P}_c}{\tilde{P}_t}.
 \end{equation}
 For nonnegligible $\zeta$ the simple connection (\ref{penguins}) 
 between $P$ and $P'$ gets modified to:
 \begin{equation}
 \tilde{P}=
 -\tilde{P}' \frac{1}{1-\zeta}
 \left( \left| \frac{V_{td}}{V_{ts}} \right| e^{-{i\beta}}
 -\lambda \zeta \right).
 \end{equation}
 Estimates of $\zeta$ using the perturbative approach of ref.\cite{BSS}
have been performed in ref.\cite{BurasFleischer1995} with the result
that
\begin{equation}
\label{limits}
0.2  {\begin{array}{c} <\\*[-12pt]\sim\end{array}} 
\left| \frac{\tilde{P}_c-\tilde{P}_u}{\tilde{P}_t-\tilde{P}_u} \right| 
{\begin{array}{c} <\\*[-12pt]\sim\end{array}} 0.5
\end{equation}
and
\begin{equation}
70^o {\begin{array}{c} <\\*[-12pt]\sim\end{array}} 
\arg {\frac{\tilde{P}_c-\tilde{P}_u}{\tilde{P}_t-\tilde{P}_u}}
{\begin{array}{c} <\\*[-12pt]\sim\end{array}} 130^o.
\end{equation}
Although the above numbers are certainly very uncertain it is 
interesting to see how the inclusion of a charmed penguin of
this size will affect the results of the fits of Section 2.
In the fit discussed here we assume that $S'$ gets modified in 
a way completely analogous to that for $P'$ (cf. Eq.(\ref{PSprim})).
As Fig. 1 shows (for which we have selected the limiting
cases of $\arg \zeta = 0$ and $180^o$), including
penguin contributions from the 
charmed-quark loops  (and assuming negligible $u$-quark terms)
may shift down the extracted value of $\gamma $ significantly.
Specifically, for $\zeta = 0.4$ the shift is of the order of $10^o$.
However, the value of $\chi ^2$ is not meaningfully smaller (Table 1). 
Furthermore, problems persist with the description of $B \to$
$\pi ^o K^o$, $\pi ^o \pi ^o$, and
$\pi ^+ \eta$ decays (Table 1).
Slightly larger values of $\zeta $ may shift $\gamma $ much more (see Fig.1).
In fact, some calculations suggest that the contributions from the charmed
penguins could be much larger than the upper limit of Eq.(\ref{limits}).
For comparison, the calculations in the second reference of \cite{Pham} 
correspond to 
$|\zeta | \approx 2$, ie. to charming penguins being dominant.

\section{Conclusions}
From the considerations of this paper it follows that:\\
(1) shifts in the extracted value of
$\gamma $, obtained in the fits with nonzero 
$\tilde{P}_c$  (and negligible $\tilde{P}_u$) 
of the size suggested by SD dynamics, are quite similar to those
found  in ref.\cite{ZL2003} for nonzero $\tilde{P}_u$
(and vanishing $\tilde{P}_c$), and \\  
(2) given the uncertainty in the size of both $u-$ and $c-$type penguins
(as well as in the strong phases of all amplitudes),
a reliable extraction of $\gamma $ requires using additional
information (data on asymmetries), possibly combined with a
judicious choice of data either insensitive or least sensitive to
such uncertainties.
This may be achieved by
restricting the considerations to the analysis of
the branching ratios and asymmetries of the
$B \to \pi K$ decays \cite{BF1999}. Clearly,
all information provided by the $B \to \pi K$ sector will be
included in the fits to all $B \to PP$ decays, if these fits take into account 
not only the branching ratios but also the asymmetries.
At present, such fits based on the branching ratios only seem to depend quite
strongly on the precise values of the latter.

This work was supported in part 
by the Polish State Committee for Scientific
Research (KBN) as a research project in 
years 2003-2006 (grant 2 P03B 046 25).

\newpage
FIGURE CAPTIONS

Fig. . Dependence of $\chi ^2$ 
on $\gamma $: a) $\tilde{P}_t$ only  - solid line ($\zeta = 0$); 
b) $\tilde{P}_t$ with corrections: long-dashed line - 
$\zeta =0.4 $; short-dashed line -
$\zeta =0.6$; dotted line -
 $\zeta =-0.6$.

\newpage

\begin{table}[t]
\caption{Fits to branching ratios of $B \to PP $ decays
(in units of $10^{-6}$) }
\label{FSIfits}

\begin{tabular}{|cc|cc|cc|}
\hline 
decay & expt & \multicolumn{2}{c|}{SD $P_t$ only } 
& \multicolumn{2}{c|}{SD $P_{t,c}$ with $\zeta = +0.4$}
 \\
&& ${\cal B}$ & deviation  
& ${\cal B}$ & deviation  \\
&&  &  (in stand. dev's) 
&  & (in stand. dev's) \\
\hline

$B^+ \to$ $ \pi ^+ \pi ^0$           & 
$5.3 \pm 0.8 $&  $4.00 $&   $1.6 $  & $4.53$  &$1.0$ \\

\phantom{$B^+ \to$} $K^+\bar{K}^0 $  &  
$ 0.0 \pm 2.4$&   $0.58 $&  $0.2$   & $0.56$   &$0.2$ \\

\phantom{$B^+ \to$} $\pi ^+\eta $    &
$4.2 \pm 0.9 $&  $2.66 $ &    $1.7$  & $2.34$   &$2.1$  \\

\phantom{$B^+ \to$} $\pi ^+\eta '$   &
$0.0 \pm 4.5 $&  $1.29$ &    $0.3$  & $1.13$    &$0.3$  \\
\hline

$B^0_d \to$ $\pi ^+ \pi ^-$          &
$ 4.6 \pm 0.4 $& $5.00$ &   $1.0$   & $4.93$   &$0.8$  \\

\phantom{$B^0_d \to$} $\pi ^0 \pi ^0$& 
$1.9 \pm 0.5 $&  $0.47$ &   $2.9$   & $0.54$   &$2.7$ \\

\phantom{$B^0_d \to$} $K^+K^-$       & 
$0.0 \pm 0.6 $ &  $0.0$ &   $0 $   & $0.0$    &$0$ \\

\phantom{$B^0_d \to$} $K^0 \bar{K}^0$&
$0.0 \pm 1.8 $ &  $0.54$ &  $0.3$   & $0.52$   &$0.3$ \\
\hline

$B^+ \to $ $\pi ^+ K^0$              & 
$21.8 \pm 1.4 $ & $21.04$ &  $0.5$  & $21.79$    &$0.0$  \\

\phantom{$B^+ \to$} $\pi ^0 K^+$     &
$12.8 \pm 1.1$ &  $12.68$ &   $0.1$  & $12.61 $  &$0.2$ \\

\phantom{$B^+ \to$} $\eta K^+$       &
$3.2 \pm 0.7 $ &  $2.53$ &   $1.0$  & $2.32 $  &$1.3$ \\

\phantom{$B^+ \to$} $\eta ' K^+$     & 
$77.6 \pm 4.6 $&  $76.44$ &   $0.3$ & $76.60$    &$0.2$  \\
\hline

$B^0_d \to$ $\pi ^- K^+$             & 
$18.2 \pm 0.8 $&  $19.00$ &  $1.0$  & $18.76$   &$0.7$  \\

\phantom{$B^0_d \to$} $\pi ^0 K^0$   &
$11.9 \pm 1.5 $&  $7.76$  &  $2.8$  & $8.02$  &$2.6$  \\

\phantom{$B^0_d \to$} $\eta K^0$     &
$0.0 \pm 4.6  $&  $2.31$  &  $0.5$   & $2.28 $  &$0.5$ \\

\phantom{$B^0_d \to$} $\eta ' K^0$   &
$65.2 \pm 6.0$ &  $70.86$ &  $0.9$  & $71.68$    &$1.1$ \\
\hline
$\chi ^2$ &&26.0&&23.7&\\
$|\bar{T}|$&&2.32&&2.47&\\
$\bar{P'}$&&-4.48&&-4.56&\\
$\bar{S'}$&&-2.29&&-2.25&\\
$\gamma _{fit} $&&$82^o$&&$73^o$&\\
\hline
\end{tabular}\phantom{xx}

\end{table}

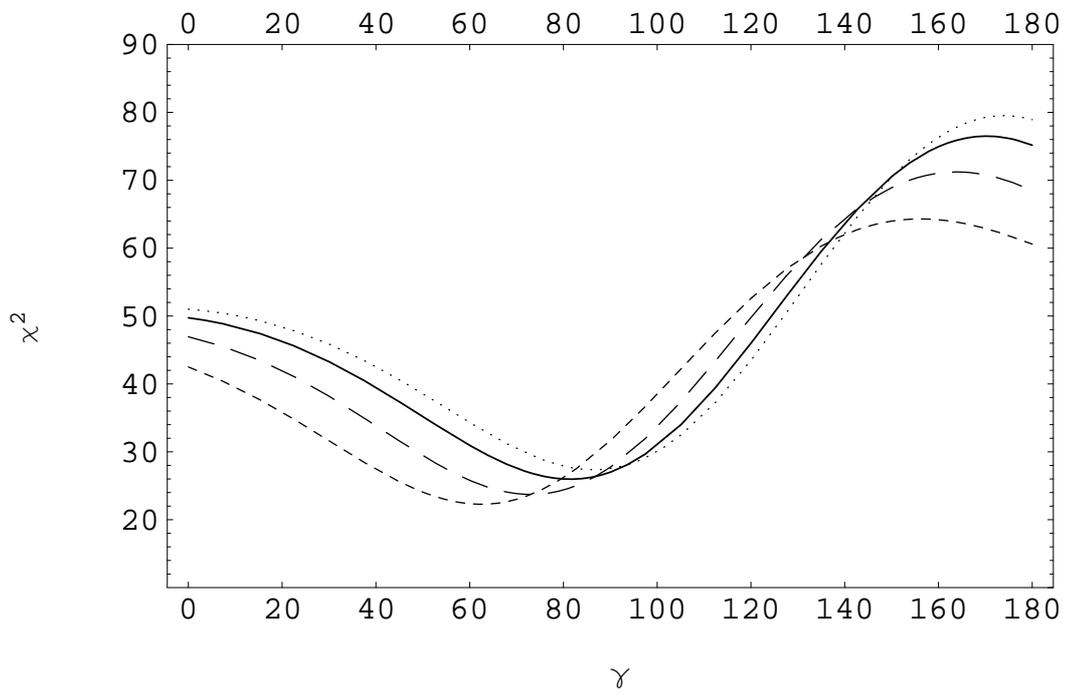
\end{document}